# A Review into eHealth Services and Therapies: Potential for Virtual Therapeutic Communities - Supporting People with Severe Personality Disorder

Dr Alice Good
University of Portsmouth
UK

Arunasalam Sambhanthan
University of Portsmouth
*UK*

**ABSTRACT**
eHealth has expanded hugely over the last fifteen years and continues to evolve, providing greater benefits for patients, health care professionals and providers alike. The technologies that support these systems have become increasingly more sophisticated and have progressed significantly from standard databases, used for patient records, to highly advanced Virtual Reality (VR) systems for the treatment of complex mental health illnesses. The scope of this paper is to initially explore e-Health, particularly in relation to technologies supporting the treatment and management of wellbeing in mental health. It then provides a case study of how technology in e-Health can lend itself to an application that could support and maintain the wellbeing of people with a severe mental illness. The case study uses Borderline Personality Disorder as an example, but could be applicable in many other areas, including depression, anxiety, addiction and PTSD. This type of application demonstrates how e-Health can empower the individuals using it but also potentially reducing the impact upon health care providers and services.

**INTRODUCTION**

ICT systems have pervaded our lives over the last couple of decades and the health care system is no exception. A relatively new buzz word, E-health is essentially an umbrella term that directly correlates to health and computing. Eysenbach (2001) explains it in the following manner:

"*e-health is an emerging field in the intersection of medical informatics, public health and business, referring to health services and information delivered or enhanced through the Internet and related technologies. In a broader sense, the term characterizes not only a technical development, but also a state-of-mind, a way of thinking, an attitude, and a commitment for networked, global thinking, to improve health care locally, regionally, and worldwide by using information and communication technology". (p. 20)*

The term arose in the late 90s when other e-words, such as e-business and e-commerce etc were also created for the purpose of portraying an increasing growth of the merging of computing into society. The convergence of health and technology has led to the development of systems and services, including: electronic health records; Health Information Systems; telemedicine; communicative wearable and portable systems, diagnostic ICT systems, well being management and monitoring.

Health Information Systems have a huge impact upon patients, health care professionals and health providers. There is the potential for patients to be more empowered as they are provided with better quality information on their conditions. They are then able to take a more pro-active approach to taking responsibility for their own well being and management of their illnesses. Health care staffs are able to communicate in a far superior way than previously, with the sharing and access of information, and advanced methods in monitoring patients. Health care providers can run a more efficient and effective service by using technology to organize resources to facilitate greater productivity.

E-health also offers considerable benefits to rural areas, particularly in developed countries. For example, telemedicine can be used to increase the accessibility of healthcare facilities to rural areas. There are however examples in less developed countries too. The issue of reaching out to rural communities, with the aim to provide specialized health care, has in recent years seen the development of a telemedicine system in Sri Lanka. The system is known as *Vidusuwa* (Sudhagar *et al.*, 2009). A study was conducted as a case study for the implementation of e-clinics in Sri Lanka, and has since evolved into a new e-business model for enhanced healthcare delivery in the developing nations. This model provided an electronic platform through which specialized consultation could be facilitated and subsequent prognosis and treatment provided. This is also an e-Solution that makes appropriate use of Electronic Medical Records (EMR) and Telemedicine technologies which enables the patient to consult a Specialist through e-Consultation. The main benefit of this solution is then enabling specialized healthcare to rural areas. There are many more examples worldwide, indicating the importance of technology in the provision of a more effective and accessible health service.

## E-HEALTH TECHNOLOGIES AND THEIR IMPACT UPON WELLBEING

E-Health lends itself well to a whole host of applications that support the wellbeing and management of mental illness, designed both from a patient and healthcare provider perspective. ICT based support is suggested as being an effective and economically viable intervention, particularly for anxiety and mood disorders (Newman *et al.*, 2011). The extent of support for people with psychiatric disorders enabled via ICTs continues to expand both in terms of development and popularity. Certainly, since the advent of Web 2.0 technologies in 2004, there has been an increase in the use of ICTs in facilitating support for people with mental health problems. Web 2.0 technologies essentially facilitate interaction between users, including social networking and collaboration. A particularly successful application of Web technologies includes 3D virtual worlds, such as Second Life (Gorini *et al.*, 2008).

Systems such as emailing, discussion forums, face book, Skype, chat rooms, VR applications and Second Life are growing in their utilization as an additional support mechanism to traditional methods (Neal & McKenzie, 2010). Research shows that both social networking sites and discussion forums have decreased the sense of 'feeling alone' for people with mental health problems (Neal & McKenzie, 2010).  Furthermore, it is proposed that there is a direct correlation between mental illness and loneliness, and that treatment for severe mental illness does not often consider the potential that social networking can have in reducing the impact of loneliness (Perese & Wolf, 2005).  Interestingly, research by Graffeo & La Barbera (2009) compared a number of online therapies, which included the use of discussion forums; chat rooms; Skype; Face Book; VR applications and Second life.  The findings suggested that both Second Life and Face book proved to be a more popular option.

The economic benefits of ICTs used in treating and supporting people with mental illness are also an important consideration. Computer and Internet assisted assessment methods and therapy programs have the potential to increase the cost-effectiveness of standardized psychotherapeutic treatments by reducing contact time with the therapist, increasing clients' participation in therapeutic activities outside the standard clinical hour, and streamline input and processing of clients' data related to their participation in therapeutic activities (Taylor & Luce 2003). In the case of second life, this becomes quite practical and even financially feasible due to the free availability of Second Life for everyone with access to the Internet.

**VIRTUAL ENVIRONMENTS SUPPORTING E-HEALTH THERAPY**

An example of ehealth therapy can be seen in VR applications. VR has developed immensely over recent years and we have seen a continuing development in the application of VR in treating medical and more recently, psychiatric conditions. VR has been used to treat a range of psychiatric conditions including Obsessive Compulsive Disorder (OCD) (Kim et al 2008), as well as phobias and Post Traumatic Stress Disorder (PTSD) (Gorrindo *et al.*, 2009). More specifically, recent research has shown how second life can be used to support people with psychiatric disorders, including the rehabilitation of people suffering with PTSD. (Anderson *et al.*, 2009; Freeman, 2008).

The application of VR as a therapeutic tool has been used in a number of mental health disorders but is particularly useful in exposure therapy. Virtual environments enable a highly visual, immersive presence in a simulated virtual world and therefore are effective in creating visual and auditory environments to treat phobias and PTSD.

**VR Applications used to treat psychosis**

VR applications have been used effectively to treat patients with schizophrenia (Freeman, 2008; Kim *et al.*, 2007). Understanding and treating schizophrenia requires consideration of patients' interactions in the social world. Withdrawal and avoidance of other people is frequent in schizophrenia, leading to isolation and rumination, and VR interactive immersive computer environments offer potential in facilitating the understanding of psychosis. Freeman (2008) used seven applications of virtual social environments to understand and treat schizophrenia namely: symptom assessment; identification of symptom markers; establishment of predictive factors; tests of putative causal factors; investigation of the differential prediction of symptoms; determination of toxic elements in the environment and development of treatment. An impairment is social skills is a recognized trait in schizophrenia and therefore the use of VR to measure social skills and perception, and then to provide training in social skills (Kim *et al.*, 2007) has potential in increasing patients sense of well being and social inclusion. The use of VR with the ascription of personalities and mental states to virtual people holds great promise in furthering the understanding and treatment of psychosis.

**VR Applications used to treat neurosis**

Research into how VR can be used to treat phobias is longstanding. Studies into using VR in treating a fear of flying were carried out as early as 1996 & 1998 (Rothbaum *et al.*, 1996; Widerhold *et al.*, 1998). Recent research reports on the application of VR to treat patients with a number of phobias including: fear of flying; fear of heights; social phobia/public speaking anxiety; spider phobia; agoraphobia; body image disturbance and obese patients (Kim *et al.*, 2008; Lynsey & Nicholas, 2007). One specific example is that of Gorrindo *et al.* (2009) who developed an exposure based therapy as an effective treatment method for patients with mental disorders such as social phobia, specific phobia and post traumatic stress disorder. This involved repeatedly presenting a patient with anxiety-producing stimuli with the goal of desensitizing a patient's anxious response. With the use of VR goggles, the patient is presented with realistic images of the specific issue. The therapist would have the control of intensity of stimulus given, so that the level of anxiety needed for the patient could be desensitized without overwhelming the patient.

A detailed example of VR applications being used to treat social phobia is portrayed in the work of Anderson *et al.* (2005). When treating social phobia, the patient is asked to read a speech in front of a virtual audience. Prior to the simulation, the therapist creates a fear-hierarchy consisting of increasingly anxiety-provoking simulations which are customized to each patients based on the degree of treatment needed. The patient is provided with a hand mounted display through which a virtual lecture hall is created. The therapist then controls the number of audience with differing response types, such as interesting, bored or even walks out. The above research conducted Anderson *et al.* (2005) shows a clear therapeutic means for treating anxiety disorder. The treatment consisted of eight individual therapy sessions, including four sessions of anxiety management training and four sessions of exposure therapy using a virtual audience, according to a standardised treatment manual. Participants are then requested to complete a standardized questionnaire assessing public – speaking anxiety at pre-treatment, post-treatment and 3 months follow up. Results showed decreases on all self-report measures of public-speaking anxiety from pre- to post-treatment, which were maintained at follow-up. This study provides preliminary evidence that a cognitive-behavioral treatment using VR for exposure to public speaking may reduce public-speaking.

The same therapy can be used to treat a fear of flying, where the patient would through a virtual airport, and wait for a boarding call. They would board an airplane and experience the take off, cruising and landing of the flight. The intensity of experience would be controlled by the therapist based on the degree of anxiety a patient has (Anderson *et al.*, 2005). .

**VR treating Post Traumatic Stress Disorder (PTSD)**

The use of VR applications in treating PTSD is particularly relevant given the reported co-occurrence of BPD and PTSD found in patients (Heffernan & Cloitre 2000; Gunderson & Sabo, 1993). Research into combat-related PTSD found that 76% of them also presented with a diagnosis of BPD (Southwick *et al.*, 1993). VR is well utilised in the treatment of PTSD (Riva, 2010; Rizzo *et al.*, 2005). The event which is known to trigger traumatic stress is created using virtual reality. One example of this type of application evaluated the effectiveness of VR therapy for a PTSD patient, who was also a survivor of September 11$^{th}$ World Trade Centre attack (Difede & Hoffman, 2002). The therapy was designed to replicate the World Trade Centre attack through VR. This featured: jets that crashed with animated explosions and sound effects, virtual people jumping to their deaths from the burning buildings, towers collapsing, and dust clouds. All contributing factors to bring the traumatic event live. The treatment was reported to be successful on reducing acute PTSD symptoms. Depression and PTSD symptoms as measured by the Beck Depression Inventory and the Clinician Administered PTSD Scale indicated a large (83%) reduction in depression, and large (90%) reduction in PTSD symptoms after completing VR exposure therapy. The results from the research suggest that VR exposure therapy shows great potential in treating acute PTSD (Difede & Hoffman, 2002).

Another example of treatment of PTSD is portrayed in a recent study known as Virtual Iraq (Geraldi *et al.*, 2008). This is a VR based therapy environment developed for treating persons from the armed forces affected by anxiety and PTSD due to their involvement in the Iraq and Afghanistan war. The idea is that a high stress environment is created. The environment features gunshots and other realistic sound effects to create a replicated war scene in the virtual world so that the patient could be exposed a realistic simulation. The therapist would control the amount of simulation through a key board. Pilot data suggest that the simulation was helpful in reducing PTSD symptoms in six of eight participants after seven sessions (Geraldi *et al.*, 2008). This could also be a possible therapeutic method which could be launched in a Second Life platform and utilised in a non lab setting. Particularly, this kind of a therapy would be economically and technologically feasible for day to day clinical practices, due to the fact it requires just basic computer applications with internet connections. An important consideration is that any VR system needs to be designed to be user friendly for therapist who may not have technical expertise (Annett & Bischof, 2010).

**An Overview of Second Life in treating and supporting people with mental illness**

Whilst VR applications are certainly developing in terms of variety and numbers in the treatment of psychiatric conditions, we are also seeing an emergence of related Second Life applications used particularly as a support tool. With approximately fifteen million users reported at the end of 2008, like Facebook, Second Life is recognized as an important meeting place for interaction (K Zero Universe, 2009).

A recent survey on healthcare related activities shows that patient education and awareness building as the major health related activity undertaken through Second Life. The second largest group of sites was that of support groups (Beard *et al.*, 2009). Research by Norris (2009) looked at the growth of healthcare support groups in virtual worlds and reported that mental health groups featured the largest number of members at 32% of the total users of Second Life. In terms of categories of groups, 15% of the health support groups in Second Life were dedicated to mental health. Second Life as a social networking medium then holds some potential in facilitating support and information sharing for people with BPD.

In terms of treatment, there are some examples of Second Life facilitating therapy. It is reported that virtual worlds can be effective in confronting phobias and addictions as well as offering potential in experimenting with new behaviors and means of expression. Research shows that Second Life has been a usual tool in facilitating exposure treatment for anxiety and behavioral activation for depression (Newman *et al.*, 2011).

**Personalized Therapy in Second Life**

The concept of personalized therapy is another area currently being researched using 3-D virtual worlds. Second life has been utilized to create a personalized-health (p-health) environment through which a personalized therapy could be provided to the patient regardless of physical and time constraints (Gorini *et al.*, 2008). The authors list the three main focus areas of p-health as follows.

- *An extended sense of presence:* P-health uses advanced simulations (3-D virtual worlds) to transform health guidelines and provisions into experience. In p-health, users do not receive abstract information but live meaningful experiences.

- *An extended sense of community (social presence):* P-health uses hybrid social interaction and dynamics of group sessions to provide each user with targeted—but also anonymous, if required—social support in both the physical and virtual world.

- *Real-time feedback between the physical and virtual worlds:* P-health uses bio and activity sensors and devices (PDAs, mobile phones) to

track both the behaviour and the health status of the user in real time and to provide targeted suggestions and guidelines. The feedback activity is twofold: (1) behaviour in physical world influences the experience in the virtual one (if I eat too much and I do not exercise, my avatar will become fatter), and (2) behaviour in the virtual world influences the experience in the real one (if I participate in the virtual support group, I can exchange SMS messages with the other participants during the day) (Gorini *et al.*, 2008).

The review highlights a number of examples of VR applications that have been designed to provide therapy for people with mental health disorders. Exposure therapies using VR are well documented, particularly in cases of PTSD. Second Life is reported as being heavily utilized in facilitating health support groups and providing health information, particularly in the area of mental health. Then there is the concept of personalized therapy and p-health environments, which could have some potential for the development of specifically tailored 3-D virtual world environments for people with mental illnesses. Finally, the economic benefit of ICT based support and therapy is also highlighted.

## 3-D VIRTUAL THERAPEUTIC COMMUNITIES: AN E-HEALTH APPLICATION FOR MANAGING MENTAL HEALTH

In this section, the paper will explore the potential for utilizing e-Health technologies in developing a 3-D virtual therapeutic community for people with mental illness and uses Borderline Personality Disorder as an example. A case study is provided of a therapeutic community used to treat Borderline Personality Disorder. The case study explores the potential of a long term cost effective model of treatment, which also empowers its users.

**Introducing Borderline Personality Disorder and support systems**

Borderline Personality Disorder (BPD) is a contentious subject, not least due to the lack of support available to patients presenting this diagnosis. People with this disorder exhibit a range of debilitating and self destructive behaviors including: depression, poor social skills and instable relationships; chemical dependency; eating disorders and suicide attempts (APA, 1994). Furthermore, approximately 60-70% people with BPD are reported to commit suicide, with approximately 10% actually being successful (Oldham, 2006, cited in NICE, 2009). The economic impact is a significant consideration with people with BPD in the UK reported as more likely to seek psychiatric intervention than people with other psychiatric disorders, and incurring primary care costs at almost twice the amount as patients with other mental illnesses (Rendu *et al.*, 2002)

Research shows that therapeutic communities (TCs) have been shown to be a valid contribution into the significant improvement into the patients' ability to cope with their negative behavioral and emotional issues (Norton, & Hinshelwood, (1996); Campling, (1999). However, the significant economic cost of treating people BPD in relation to other mental illnesses poses difficulties in providing adequate treatment via TCs (NCCMH, 2009). In addition, people with BPD often require high levels of support, which can result in emergency hospital admissions (NCCMH, 2009). The over-riding issue is on providing adequate support for people with BPD and whether there is potential for technology enabled support. Whilst the advent of social networking has given rise to online support groups for people with BPD, there is potential scope to utilise resources such as virtual, 3D environments in providing further support. This could also have some impact upon reducing emergency hospital admissions, enabling an increased sense of well being for the individual with BPD, as well as a possibility for a decrease in the economic impact to the health services.

**Therapeutic Community Model, a cost effective treatment**

The concept of therapeutic communities first arose from the treatment of veterans suffering with PTSD from the Second World War (Main, 1946). The Henderson Hospital that emerged in 1947, and recently closed due to lack of funding, is an example of a TC featuring four specific themes: democratisation, permissiveness, reality confrontation and communalism (Rapoport, 1960). TCs essentially incorporate a democratic ethos and are run with less formal roles from the professionals, with both members and professionals having an equal say in the organizations and management (Norton, K (1992). Research into the effectiveness of TCs, show that they are a viable and productive method for treating BPD (Dolan et al, 1992; Campling, P., 1999). They have been shown to be more effective and accepted than traditional hospitals, due to the lack of strict treatment regimes (NICE, 2009).

In assessing the impact of TC based treatment on core personality disorder symptoms, research shows that in using the Borderline Syndrome Index (BSI), there was a significant reduction in symptoms one year following discharge (Dolan et al, 1997). Studies by Chiesa *et al.* (1996), Davies *et al.* (1999) and Dolan *et al.* (1996), cited in Davies (2003), and looked at the economic benefits of TCs. The results showed a reduction in primary care costs, one year following treatment as well as a decrease in the utilization of services (Davies, 2003). Furthermore, the actual cost of treating people with BPD within a TC setting could be recovered by the state within two years (Dolan *et al.*, 1996). TCs could then be viewed as an effective method for treating BPD as well as reducing economic impact in the reduction of services required following successful treatment.

**User-led informal networks of care in therapeutic community**

In terms of e-Health technologies that facilitate support specifically for people with BPD, there are a number of online resources including 'BPD Today' and various support groups facilitated by Yahoo Groups and Face book. These also include networks that exemplify the principles of therapeutic communities. A recent work of Rigby & Ashman (2008) proposed service user-led informal networks of care in therapeutic community (TC) practice as a cost effective solution for BPD, which is focused on rural communities. The network features a well-established internet messaging and chat room facilities uniquely structured and moderated to encompass therapeutic community principles and provide equality of access across a huge mixed urban and rural catchment area. Both hardware and software are inexpensive, easily transferable to similar services and could be modified to suit other applications. The system has developed to provide out-of-hours support and to extend the work of a number of day therapeutic communities for service users with moderate to severe personality disorders.

TCs have been shown to be an effective method of treatment and already, we are seeing ICT, user-led networks of patients that incorporate TC principles, though this is limited to emails and discussion forums. There is huge potential for utilizing technologies such as Second Life in hosting a virtual TC specifically for people with BPD. The review earlier demonstrated how popular Second Life is for enabling social interaction within health support groups, particularly mental health. Second Life would certainly enable an enhanced sense of immersion and presence as opposed to static methods of communication and interaction.

**DISCUSSION**

Whilst the scope for a support group for people BPD in Second Life seems likely, there is further potential to develop an environment that incorporated TC principles. Furthermore, given the reported effectiveness of VR applications in treating PTSD and other mental health problems, there could be potential in treating some of the traits of BPD, including reality confrontation of debilitating behaviors.

Such a proposal will undoubtedly require the collaboration of specialized professionals and the co-operation of outreach services, as well as the input of end users. In terms of the design process, an end-user centered approach would be applied where clients presenting a diagnosis of BPD contribute to the viability of this work, along with professionals specializing in the treatment of this condition. A triangulation approach will be applied in the requirements stage, using a focus group and interviews to validate the results and to provide a deeper understanding.

**A Model of User-Led Networks**

The technology-based user-led network system in supporting people with BPD in rural communities (Rigby & Ashman, 2008) exemplifies principles of effective treatment via TCs. The proposal of a therapeutic community facilitated in Second Life would seek to draw on the principles advocated here:
- To provide greater containment of anxiety and to avoid total hospitalization.
- To extend therapeutic community principles outside the working day allowing more opportunities for new learning and real-life opportunities to take responsibility for self and others.
- To promote attachment to, and exchange within a group, both to avoid the repetition of unhelpful dynamics between individuals and to practice appropriate help-seeking behaviors.

- To reduce the burden of care on the professional network beyond the therapeutic community (Rigby & Ashman, 2008).

**Specific Needs Pertaining to People with BPD**

There are also the inherent needs of people with BPD that should be addressed when considering the design of a Virtual TC.
- People with BPD have a strong need to feel accepted, heard and understood.
- They need a sense of safety; for instance consistent people and places where they can become attached while working through their difficulties.
- People with BPD often require an instant response when in crisis, whether it is in the day or at night (NICE, 2009).

**Identified Benefits of a virtual TC**

A virtual TC, hosted in Second Life could be considered to be beneficial to people with BPD for the following reasons:
- An extended sense of presence as opposed to static discussion forums.
- Anonymity through the use of avatars.
- An extended sense of community and sharing experiences and support.
- A sense of ownership, in accordance with ethos of traditional TCs.
- Reducing the sense of feeling alone.
- Potential for exposure therapy in reality confrontation.
- 24/7 access to support.
- Economic benefit to Primary Services.

One of the challenges in treating people with BPD is that although they can often present as being in crisis, it can prove to be difficult to engage them in any treatment (Winston, 2000). A virtual TC, because of its ease of access and potential anonymity, could then be a viable system of treatment and support.

**RECOMENDATIONS**

The development of a 3-D virtual therapeutic community would require further research in assessing the requirements of. A broad explanation of the requirements follows:

- An evaluation of existing methods of ICT enabled support for people with BPD to inform the proposal and to assess how to incorporate elements of good practice. This will include social networking sites and the virtual TC system developed by Rigby and Ashman (2008).

- A focus group of professionals working with people with BPD will facilitated to assess how best to design and promote the proposed support system

- Qualitative and quantitative data derived from surveys and interviews with both patients diagnosed with BPD and related professionals, in understanding their needs and requirements.

- Collaboration with two consultant psychiatrists in psychotherapy, specializing in the treatment BPD.

- A user centered approach to the design of a therapeutic community is Second Life will be considered mandatory. Human Computer Interaction (HCI) principles will be applied through all stages of design. User involvement will feature from requirements through to testing.

**Summary & Conclusions**

The use of ICT based support and treatment of mental disorders continues to expand as increasingly more people are seeking virtual support and/or therapies. The use of ICT based support specifically for people with BPD however, is limited predominantly to discussion forums, emails and social networking sites facilitated by Face Book. The review indicates that there are several interesting VR based therapies in place to treat patients with mental health problems. In addition, the popularity of Second Life in facilitating health information sharing and support groups is also highlighted. There is however neither known VR therapies nor Second Life sites that are designed specifically for people with BPD.

This paper has highlighted the significant contribution that TCs have in the treatment, support and rehabilitation of people with BPD (Norton, K. & Hinshelwood, R. D., 1996; Campling, P., 1999). It also portrays the ongoing work of Rigby and Ashman (2008) where technology is used to facilitate TC networks, specifically for people with BPD, in rural communities. The paper also highlights the significant economic impact upon primary care resources that people with BPD place, in comparison to those exhibiting other mental health disorders.

There is certainly great potential for using 3-D VR environments such as Second Life to host a TC specifically tailored for people with BPD. Examples include: reducing the sense of feeling alone; shared support network; containment; resources on self help techniques and of course empowering patients. In addition, there is scope for incorporating specific therapies too, with the input of professionals. Such a system could also be economically viable, particularly in comparison with traditional primary costs, as well as reducing the burden on health services. To ensure that the proposed virtual TC is well utilised, there needs to be a clear match between the users' needs and the design of such a system. A user centered approach to the design, which involves all potential groups of users will help to ensure that a virtual TC hosted in Second Life will be usable and accepted. The recommendations in this paper will certainly go some way in assisting this process. Hence, the application of Second life based therapeutic communities to treat BPD patients could be an innovative area of focus in providing a new avenue for further research development in the field of treating and supporting people with BPD.

Further research is of course necessary to assess the likely acceptance of this type of support and how well it will be utilsed. The fact that Second Life has been shown to be a very popular medium in hosting support groups, particularly in the area of mental health (Norris, 2009) lend to the viability of such a proposal. Furthermore, it is a very low cost and accessible resource.

Linehan,M.M., Schmidt,H., Dimeff,L.A., Craft,J.C., Kanter,J., Comtois,K.A. (1999). Dialectical behavior therapy for patients with borderline personality disorder and drug-dependence. *American Journal on Addiction*, 8(4), 279-292.

Main T. (1946). "The Hospital as a Therapeutic Institution". *Bulletin of the Menninger Clinic* 10: 66–70.

Neal, D.M., & McKenzie, P.J. (2010) "I did not realize so many options are available": Cognitive authority, emerging adults, and e- mental health, *Library & Information Science Research* (2010).

Newman G.M., Szkodny E.L., Llera J.S. and Przeworski A. (2011), *A review of technology-assisted self-help and minimal contact therapies for anxiety and depression: Is human contact necessary for therapeutic efficacy?* Clinical Psychology Review. Volume 31, Issue 1, February 2011, Pages 89-103.

NICE (2009) *Borderline Personality Disorder – The NICE guideline on treatment and management*. NICE, Royal College of Psychiatrists and the British Psychological Society.

Norris, J., (2009). The Growth and Direction of Healthcare Support Groups in Virtual Worlds. 3D Virtual Worlds for Health and Healthcare. Volume 2, Number 2.

Norton, K. (1997). In the Prison of Severe Personality Disorder, *Journal of Forensic Psychiatry & Psychology*, vol. 8, issue. 2, pp. 285 - 298.

Norton, K (1992). Personality disordered individuals: the Henderson Hospital model of treatment. *Criminal behaviour and mental health*, Vol 2: 180-191Norton, K. & Hinshelwood, R. D. (1996) Severe personality disorder: treatment issues and selection for in-patient psychotherapy. *British Journal of Psychiatry*, **168**, 723-731.

Perese, E., & Wolf, M. (2005). Combating loneliness among persons with severe mental illness: Social network interventions' characteristics, effectiveness, and applicability. *Issues in Mental Health Nursing*, 26(6), 591-609.

Rapoport, R (1960*). Community as doctor'*. Tavistock: London.

Rigby, M. & Ashman, D. (2008), Service Innovation: A Virtual informal network of care to support a 'lean' therapeutic community in a new rural personality disorder service, *The Psychiatrist*, vol. 32, pp. 64 - 67.

Riva, G., Raspelli, S., Algeri, D., Pallavicini, F., Gorini, A. Wiedrhold, K. & Gaggioli, A (2010), Inter reality in Practice: Bridging Virtual and Real Worlds in the Treatment of Posttraumatic Stress Disorders, *Cyber Psychology & Behavior*, vol. 13.

**KEY TERMS AND DEFINITIONS**
eHealth; Second Life; Virtual Reality; ICTs; Therapeutic Communities.

**eHealth:** Transfer of Health by Electronic Means
**Second Life:** Forum for Online Interactive Virtual Worlds
**Virtual Reality:** 3D simulated interactive environment
**ICTs:** Information Communication Technologies
**Therapeutic Communities:** Participatory, Group Approach to Therapy